\documentstyle[12pt]{article}
\title{On the complex structure in the Gupta--Bleuler quantization
method}
\author{S. Bellucci\thanks{e-mail: bellucci@lnf.infn.it}\\
INFN-Laboratori Nazionali di Frascati, \\
P.O. Box 13, 00044 Frascati, Italy\\
 and \\
A. Galajinsky\thanks{e-mail: galajin@phys.tsu.tomsk.su}\\
Department of Theoretical Physics, \\
Tomsk State University, 634050 Tomsk, \\
Russian Federation}
\date{}
\begin{document}
\maketitle
\large
\begin{abstract}
We examine the general conditions for the existence of the complex
structure intrinsic in the Gupta-Bleuler quantization method
for the specific case of mixed first and second class fermionic constraints
in an arbitrary space-time dimension. The cases $d=3$ and 10 are shown
to be of prime importance. The explicit solution for $d=10$ is presented.
\end{abstract}

\section{Introduction}

Since its invention in quantum electrodynamics [1], the Gupta-Bleuler
method has become a conventional tool when quantizing theories
with anomalies and/or second class constraints. In the latter case, it
requires the construction of a specific complex structure $J$ on a
phase space of a model which allows one to split the original second class
constraints into (complex conjugate) holomorphic and antiholomorphic
sets [2,3]. The existence of such a $J$ in a neighborhood of a (second
class) constraint surface has been proven in Ref. [4]. It was stressed
in [4], however, that, generally, this may break manifest covariance in
a problem. If the second class constraints are a-priori in the  holomorphic
representation, the Gupta-Bleuler method was shown to admit an elegant
BRST formulation [2,4--6], which involves a pair of (Hermitean
conjugate) BRST charges (see also related works [7,8]).

As it has already been discussed in Ref. [2], the approach
applies also to the specific case of mixed first and second class
fermionic constraints  $L_\alpha\approx0$ \begin{equation}
\begin{array}{l}
\{L_\alpha,L_\beta\}=2i(\Gamma^np_n)_{\alpha\beta}\equiv
\Delta_{\alpha\beta}, \qquad \alpha=1,\dots,n,\\
\{L_\alpha,p^2\}=0, \\
{\rm rank}\,(\Gamma^np_n)_{\alpha\beta}=n/2,\\
\end{array}
\end{equation}
with $p^n$ a light-like vector $p^2=0$, which is just the case when
studying the superparticle [9] and superstring [10] theories in flat
superspace \footnote{For simplicity, in what follows we shall discuss
the superparticle case only.}. Following the procedure, one has to
extract first class constraints from the original mixed system
$L_\alpha\approx0$ in the covariant (and reducible) way
\begin{equation}
(\Gamma^np_nL)_\alpha=0.
\end{equation}
and then solve the equations
\addtocounter{equation}{1}
$$
{J_\alpha}^\beta {J_\beta}^\gamma=-{\delta_\alpha}^\gamma,
\eqno{(3a)}$$
$$
{J_\alpha}^\beta\Delta_{\beta\gamma}+\Delta_{\alpha\beta}
{J_\gamma}^\beta=0,
\eqno{(3b)}$$
$$
\{L_{\alpha},{J_{\beta}}^\gamma\}=0, \qquad
\{{J_\alpha}^\beta\ ,{J_\gamma}^\delta\}=0
\eqno{(3c)}$$
$$
\{{J_\alpha}^\beta,p^2\}=0,
\eqno{(3d)}$$
for the complex structure. With $J$ at hand, the mixed constraints can
be split into holomorphic
${L^-}_\alpha\equiv {p^-}_\alpha{}^\beta L_\beta\approx0$
and antiholomorphic ${L^-}_\alpha\equiv {p^-}_\alpha{}^\beta L_\beta\approx0$
sets
\begin{equation}
\begin{array}{l}
\{{L^-}_\alpha,{L^-}_\beta\}\approx0, \qquad
\{{L^+}_\alpha,{L^+}_\beta\}\approx0,\\
\{{L^+}_\alpha,{L^-}_\beta\}\approx{p^+}_\alpha{}^\gamma
\Delta_{\gamma\beta},\\
({L^+}_\alpha)^*={L^-}_\alpha,\end{array}
\end{equation}
where ${p^\pm}_\alpha{}^\beta\equiv \displaystyle\frac 12
({\delta_\alpha}^\beta\pm i{J_\alpha}^\beta)$. Half of these can
further be used to define physical states in a complete Hilbert space
\begin{equation}
\hat L^-{}_\alpha|{\rm phys}\rangle=0, \qquad
(\Gamma^n\hat p_n\hat L)_\alpha|{\rm phys}\rangle=0, \qquad
\hat p^2 |{\rm phys}\rangle=0. \qquad
\end{equation}
Note that only half of the original second class constraints were
effectively used in Eq. (5).

The scheme outlined was shown to admit a remarkably simple
solution for the $4d$ superparticle [5]
\begin{equation}
J\sim \gamma^5.
\end{equation}
In that case, Eq. (5) was proven to produce a massless irreducible
representation of the super Poincar\'e group of superhelicity 0 (on-shell
massless chiral scalar superfield).

However, the general conditions for the existence of $J$ in an
arbitrary space-time dimension are unknown yet. It is this problem
which we address in the present work. As shown below, the cases $d=3$ and 10
are of prime interest. The explicit solution for $10d$ is the main
result of this letter.

In the next section we briefly outline our strategy and specify the
dependence of the formalism on the dimension of space-time. Section 3
contains a solution for $d=10$ in the Hamiltonian formalism. As shown
below, the possibility to construct a covariant $J$ satisfying Eq. (3)
requires an extension of the original phase space by additional
(unphysical) variables. In Sec. 4 we construct the corresponding $10d$
Lagrangian formulation. Some comments on the problem in $3d$ superspace
are presented in Sec. 5. In particular, we show that the Gupta-Bleuler
approach fails in that case. We end the paper with the discussion of
open problems in Sec. 6. Some technical details are gathered in
Appendix.

\section{How to attack the problem?}

In short, our proposal is to decompose the tensor ${J_\alpha}^\beta$
into irreducible representations of the Lorentz group and then reduce
Eq. (3) to equations for those irreps. It should be stressed, that the
structure of the equations will essentially depend on the dimension of
space-time and the type of spinors existing in a given dimension.
If a spinor entering into the superparticle action is of Dirac type,
the corresponding constraints will automatically belong to the
holomorphic representation. In $d=2,3,4$ (mod $8$) a spinor can be chosen
to be Majorana (and Majorana-Weyl in $d=2$ (mod $8$)). For even
dimensions it can be decomposed into the direct sum of two (complex
conjugate) Weyl spinors, which, again, brings the constraints to the
holomorphic representation. Thus, if by analogy with the superstring
theory one restricts ourselves with the case $d<11$, there remain only
two physically nontrivial possibilities $d=3$ and 10 which we examine
below. Note also, that, generally, the possibility to realize a
covariant $J$ satisfying Eq. (3) will require an extension of the
original phase space by additional (unphysical) variables.

\section{A solution for $d=10$}

A minimal spinor representation of the Lorentz group in $d=10$ is
realized on Majorana--Weyl spinors (16 real components). Chiral spinors
are distinguished by the position of their indices $\psi^\alpha$,
$\varphi_\alpha$, $\alpha=1,\dots,16$ (for the $10d$ spinor notation
see Ref. [11]). By making use of the Fierz identity\footnote{In what
follows, $\Gamma^{ab},\Gamma^{abcd},\dots$ denote the totally
antisymmetrized product of the $10d$ $\Gamma$-matrices (see also
Appendix).  } [12]
\begin{equation}
32{\delta_\alpha}^\delta{\delta_\gamma}^\beta=2{\delta_\alpha}^\beta
{\delta_\gamma}^\delta-\Gamma^{ab}{}_\alpha{}^\beta
\Gamma_{ab}{}_\gamma{}^\delta+\frac 2{4!}\Gamma^{abcd}{}_\alpha{}^\beta
\Gamma_{abcd}{}_\gamma{}^\delta
\end{equation}
one can decompose ${J_\alpha}^\beta$ into irreducible pieces
\begin{equation}
{J_\alpha}^\beta= {\delta_\alpha}^\beta J +\Gamma^{ab}{}_\alpha{}^\beta
J_{ab}+\Gamma^{abcd}{}_\alpha{}^\beta J_{abcd}.
\end{equation}
A substitution of this into Eq. (3b) yields (see also Appendix)
\begin{equation}
2\Gamma^n(Jp_n+2J_{nm}p^m)+2\Gamma^{abcdn}J_{[abcd}p_{n]}=0,
\end{equation}
or
\begin{equation}
Jp_n+2J_{nm}p^m=0, \qquad J_{[abcd}p_{n]}=0.
\end{equation}
Analogously, Eq. (3a) reads
\addtocounter{equation}{1}
\begin{eqnarray*}
&& 1+J^2-2J_{ab}J^{ab}+b_3J_{abcd}J^{abcd}=0, \qquad \qquad \\
&& 2JJ_{ab}+a_2J_{abcd}J^{cd}+\displaystyle\frac 1{2!}
\epsilon_{abcdefghij}J^{[cdef}J^{ghij]}=0, \qquad
\qquad \qquad \qquad (11)\\
&& 2JJ_{abcd}+J_{[ab}J_{cd]}+a_1J_{[abc\hat m}{J^m}_{d]}
+b_2J_{[ab\hat m\hat n}J^{mn}{}_{cd]}-\\
&& -\displaystyle\frac 1{4!}\epsilon_{abcdefghij}\Big(J^{[ef}J^{ghij]}+
b_1J^{[efg\hat m}{J_m}^{hij]}\Big)=0, \qquad \\
\end{eqnarray*}
with $a_1$, $a_2$, $b_1$, $b_2$, $b_3$ denoting some constants (in what
follows we will not need their explicit form) and
$\epsilon_{abcdefghij}$ the $10d$ Levi-Civita tensor.

In obtaining Eq. (11) the identities [12]
\begin{equation}
\begin{array}{l}
\Gamma^{(n)a_1\dots a_n}{}_\alpha{}^\beta=(-1)^{n/2}\displaystyle
\frac{\epsilon^{a_1\dots a_na_{n+1}\dots a_{10}}}{(10-n)!}
\Gamma^{(10-n)}{}_{a_{n+1}\dots a_{10}}{}_\alpha{}^\beta\qquad
\mbox{$n$ even}\\
\Gamma^{(n)a_1\dots a_n}{}_{\alpha\beta}=(-1)^{(n-1)/2}\displaystyle
\frac{\epsilon^{a_1\dots a_na_{n+1}\dots a_{10}}}{(10-n)!}
\Gamma^{(10-n)}{}_{a_{n+1}\dots a_{10}\,\alpha\beta}\qquad
\mbox{$n$ odd}\end{array}
\end{equation}
have been used.

In the presence of the tensor $J_{abcd}$ Eq. (11) looks rather
complicated. It is instructive then to try setting
\begin{equation}
J_{abcd}=0,
\end{equation}
which considerably simplifies Eqs. (10) and (11)
\addtocounter{equation}{1}
$$
Jp_n+2J_{nm}p^m=0,
\eqno{(14a)}$$
$$
1+J^2-2J_{ab}J^{ab}=0,
\eqno{(14b)}$$
$$
JJ_{ab}=0,
\eqno{(14c)}$$
$$
J_{ab}J_{cd}+J_{ac}J_{db}+J_{ad}J_{bc}=0.
\eqno{(14d)}$$
It turns out that this system does admit a solution. Actually, from Eq.
(14c) it follows that either $J=0$ or $J_{ab}=0$ (the choice when they
are both equal to zero is in a contradiction with Eq. (14b)). It is
straightforward to check that the latter case leads to a contradiction
between Eq. (14a) and (14b) (it is enough to consider those equations
in the rest frame $p_n=(E,0,\dots,0,E)$). Thus, one has to put
\begin{equation} J=0,
\end{equation}
which brings Eq. (14) to the form
\addtocounter{equation}{1} $$ J_{nm}p^m=0, \eqno{(16a)}$$ $$
1-2J_{ab}J^{ab}=0,
\eqno{(16b)}$$
$$
J_{ab}J_{cd}+J_{ac}J_{db}+J_{ad}J_{bc}=0.
\eqno{(16c)}$$
The simplest solution to Eq. (16c) reads
\begin{equation}
J_{ab}=\frac 1\alpha (A_aB_b-A_bB_a),
\end{equation}
with $A,B$ denoting some vectors and $\alpha$ a scalar. The
substitution of this into Eq. (16b) determines $\alpha$:
\begin{equation}
\alpha=\pm2\sqrt{A^2B^2-(AB)^2}.
\end{equation}
In addition, Eq. (16a) requires
\begin{equation}
(Ap)=0, \qquad (Bp)=0,
\end{equation}
because, otherwise, it would mean that $A$ and $B$ are linearly
dependent and, hence, $J_{ab}=0$. One can check, further, that it is
impossible to construct two vectors $A,B$ (satisfying Eq. (19)
on the constraint surface) from the phase space variables of the
original superparticle model. This suggests an extension of the space
by two new variables $A^n$, $B^n$. In order for them to be
nondynamical, they should be subject to the (first class) constraints
\begin{equation}
p_{A\,n}=0, \qquad p_{B\,n}=0,
\end{equation}
In Eq. (20) $p_A,p_B$ denote momenta canonically conjugated to $A,B$
respectively. Following this course, Eq. (19) can further be
incorporated into the scheme as gauge fixing conditions for some of the
constraints (20). Actually, let us extend the original phase space by
one more canonical pair $(\Lambda^n, P_{\Lambda\,n})$. In order to
suppress the dynamics in this sector, we impose the constraints [11]
\begin{equation} \begin{array}{c}
\Lambda^2=0, \qquad (\Lambda p)=-1,\\
{P_\Lambda}^n=0,\end{array}
\end{equation}
or, equivalently,
\addtocounter{equation}{4}
$$
\Lambda^2=0, \qquad (P_\Lambda p)=0,
\eqno{(22a)}$$
$$
(\Lambda p)=-1, \qquad (P_\Lambda\Lambda)=0,
\eqno{(22b)}$$
$$
P_{\Lambda\,n}+\Lambda_n(P_\Lambda p)+p_n(P_\Lambda\Lambda)=0.
\eqno{(22c)}$$
The constraints (22a), (22b) are second class, while there are only
eight linearly independent first class ones in Eq. (22c) (one can find
two identities for the constraint set (22)). Note that the total number
of constraints (22) is sufficient to suppress just one canonical
pair of variables.

Let us now impose two gauge fixing conditions to the first class
constraints (20)
$$ p_{A\,n}=0, \qquad (Ap)=0, \eqno{(23a)}$$ $$
p_{B\,n}=0, \qquad (Bp)=0,
\eqno{(23b)}$$
and make use of $\Lambda$ to split these into first and second class
$$
(Ap)=0, \qquad (p_A\Lambda)=0,
\eqno{(24a)}$$
$$
p_{A\,n}+p_n(p_A\Lambda)=0,
\eqno{(24b)}$$
$$
(Bp)=0, \qquad (p_B\Lambda)=0,
\eqno{(25a)}$$
$$
p_{B\,n}+p_n(p_B\Lambda)=0.
\eqno{(25b)}$$
The constraints (24a) ((25a)) are second class, whereas the first class
ones (24b) ((25b)) contain only nine linearly independent
components. The total number of constraints is sufficient to suppress
the dynamics in the sector $(A,p_A)$, $(B,p_B)$. In other words, the
variables $(A,p_A)$, $(B,p_B)$ are unphysical. Note also, that Eqs. (3c)
and (3d) automatically holds when extending the space in this way
\footnote {In passing a to quantum description one has to fix completely the
gauge freedom in the sector $(A,p_A)$, $(B,p_B)$ because, otherwise, the
vanishing of the first class constraints (24b) and (25b) on physical
states would be incompatible with the prescription (5). In order to maintain
the manifest covariance in the problem, it is sufficient to introduce eight
sectors like Eq. (21), when imposing a gauge choice. The corresponding
Lagrangian formulation will be presented elsewhere.}.

Thus, in the enlarged phase space Eqs. (8),(13),(15),(17) and (18)
realize the needed complex structure.

\section{A $10d$ Lagrangian formulation}

A Lagrangian formulation which reproduces Eqs. (21) and (23) when
passing to the Hamiltonian formalism reads
\begin{eqnarray}
S=\displaystyle\int d\tau \frac 1{2e} (\dot x^n-i\theta\Gamma^n
\dot\theta-\omega\Lambda^n-\omega_1A^n-\omega_2B^n)^2-\omega-\Phi\Lambda^2.
\end{eqnarray}

As compared to the Casalbuoni-Brink--Schwarz model, this Lagrangian
involves a set of auxiliary variables
($\omega,\omega_1,\omega_2,\Phi,\Lambda^n, A^n,B^n$).

Moving to the Hamiltonian formalism one finds the primary constraints
(we denote as ($p_e,p,p_\theta, p_\omega,
p_{\omega_1},p_{\omega_2},p_\Phi, p_\Lambda,p_A,p_B$) the
momenta canonically conjugated to the variables
($e,x,\theta,\omega,\omega_1,
\omega_2,\Phi,\Lambda, A,B$), respectively)
\begin{equation}
\begin{array}{l}
p_e=0, \qquad
p_\theta+i\theta\Gamma^np_n=0, \qquad p_\Phi=0,\\
p_\omega=0, \qquad p_{\omega_1}=0, \qquad p_{\omega_2}=0, \\
{p_\Lambda}^n=0, \qquad {p_A}^n=0, \qquad {p_B}^n=0,
\end{array}
\end{equation}
and the
relation determining $\dot x^n$ as a function of some of the remaining
variables \begin{equation} \dot
x_n=ep_n+i\theta\Gamma_n\dot\theta+\omega\Lambda_n+
\omega_1A_n+\omega_2B_n.
\end{equation}
The canonical Hamiltonian looks like
\begin{eqnarray}
&& H=(p_\theta+i\theta\Gamma^np_n)\lambda_\theta+p_e\lambda_e+
p_\omega\lambda_\omega+p_{\omega_1}\lambda_{\omega_1}+
p_{\omega_2}\lambda_{\omega_2}+p_\Phi\lambda_\Phi
+p_\Lambda\lambda_\Lambda\cr
&&+p_A\lambda_A+p_B\lambda_B+
e\displaystyle\frac{p^2}2+\omega(1+\Lambda p)+\omega_1(pA)+
\omega_2(pB)+\Phi\Lambda^2,
\end{eqnarray}
where the $\lambda's$ denote Lagrange multipliers enforcing the
primary constraints.

The preservation in time of the primary constraints implies the secondary
ones
\addtocounter{equation}{1}
$$
\Lambda^2=0, \qquad \Lambda p+1=0, \qquad p^2=0,
\eqno{(30a)}$$
$$
(pA)=0, \qquad (pB)=0,
\eqno{(30b)}$$
$$
\omega_1 p^n=0, \qquad \omega_2 p^n=0,
\eqno{(30c)}$$
$$
\omega p^n+2\Phi\Lambda^n=0,
\eqno{(30d)}$$
and determines half of the $\lambda_\theta$
\begin{equation}
\Gamma^np_n\lambda_\theta=0.
\end{equation}
Consider now Eq. (30d). Multiplying it by $\Lambda^n$, $p^n$ and
taking into account Eq. (30a) one gets
\begin{equation}
\omega=0, \qquad \Phi=0.
\end{equation}
Analogously, Eq. (30c) provides us with
\begin{equation}
\omega_1=0, \qquad \omega_2=0.
\end{equation}
The consistency check for the secondary constraints yields
\begin{equation}
\begin{array}{ll}
p\lambda_\Lambda=0, \qquad & \Lambda\lambda_\Lambda=0,\\
p\lambda_A=0, \qquad & p\lambda_B=0,\\
\lambda_\omega=0, \qquad & \lambda_\Phi=0,\\
\lambda_{\omega_1}=0, \qquad & \lambda_{\omega_2}=0, \\
\end{array}
\end{equation}
and no tertiary constraints appear.

Thus, the complete constraint system can be written in the form
\addtocounter{equation}{1}
$$
p_e=0, \qquad p_\theta+i\theta\Gamma^np_n=0, \qquad p^2=0,
\eqno{(35a)}$$
$$
\begin{array}{ll} \omega=0, \qquad & p_\omega=0,\\
\omega_1=0, \qquad & p_{\omega_1}=0,\\
\omega_2=0, \qquad & p_{\omega_2}=0,\\
\Phi=0, \qquad & p_\Phi=0,\\
\end{array}
\eqno{(35b)}$$
$$
\Lambda^2=0, \qquad (\Lambda p)+1=0, \qquad {p_\Lambda}^n=0,
\eqno{(35c)}$$
$$
(pA)=0, \qquad {p_A}^n=0,
\eqno{(35d)}$$
$$
(pB)=0, \qquad {p_B}^n=0.
\eqno{(35e)}$$
The constraints (35a) are just those of the Casalbuoni-Brink--Schwarz
superparticle. The constraints (35b) are second class and can be
omitted after introducing the associated Dirac bracket, which leaves
us with the needed constraints (21), (23) in Eqs. (35c)-(35e).

It is straightforward to check, further, that in the light-cone
gauge \footnote {Here we assume that $p^i\ne0$.}
\begin{equation}
\begin{array}{ll} \Gamma^+\theta=0 ,\qquad & x^+=\tau p^+,\\
e=1, \qquad & \Lambda^i=0,\\
A^2=1, \qquad & A^i=0,\\
B^2=1, \qquad & B^i=0,\\
\end{array}
\end{equation}
the physical sector of the theory described by the action (26)
coincides with the one of the
Casalbuoni-Brink--Schwarz superparticle [9]. This proves the physical
equivalence of the models.

\section{A comment on the problem in $3d$ superspace}

The minimal spinor representation of the Lorentz group in
${R}^{3/2}$ superspace is realized on Majorana spinors $\theta^\alpha$,
$\alpha=1,2$, $(\theta^\alpha)^*=\theta^\alpha$. In that case, the
fermionic constraints $L_\alpha\approx0$, $\alpha=1,2$ involve {\it
only one} second class constraint which, evidently, can not be
separated into holomorphic and antiholomorphic sets. It is
straightforward to check \footnote{To see this,
it is sufficient to decompose the
$J_{\alpha\beta}$ with respect to the complete basis $\{{\bf
1},(\Gamma^n)^T\}$ in the space of $2\times2$ complex matrices
$J_{\alpha\beta}=J\delta_{\alpha\beta}+J_n{\Gamma^n}_{\beta\alpha}$ and
plug this into Eqs. (3a), (3b).} that Eqs. (3), being reduced to the
$3d$ superspace (with  $\Delta=2iC\Gamma^n p_n$ and $C$
the charge conjugation matrix), imply
\begin{equation}
p_n=0,
\end{equation}
and, hence, possesses no physically sensible solution.

Thus, the $3d$ superparticle yields an example when the Gupta-Bleuler
approach fails.

\section{Discussion}

We conclude with some remarks and open problems.

a) When analyzing the problem in $10d$ superspace the condition
$J_{abcd}=0$ has been chosen by hands. It still remains to be
understood, whether there are other solutions to Eq. (3) and, if it is
the case, how they are related to each other.

b) (A comment on quantization) The complex structure
constructed involves $\frac 1{\sqrt{A^2B^2-(AB)^2}}$
which, generally, leads to a nonlocal operator at the quantum level.
Note, however, that this can be avoided in the covariant gauge $A^2=1,
B^2=1, (AB)=0$. Since the commutation relations in the sector
$(x,p),(\theta,p_\theta)$ are canonical, it suffices to realize
the quantum brackets in the sector
$(A,p_A),(B,p_B),(\Lambda,p_{\Lambda})$.  This work is currently in progress.

c) There are two comments on a possible generalization to superstrings.
First of all, since the super Virasoro constraints weakly commute with the
fermionic ones, Eqs. (1)-(3) will take a more complicated form.
Secondly, in a recent work [13] Berkovits proved that the
generalization of Eq.  (21) to the superstring case, which was
previously used in [11,14,15], is not harmless.  There remains one
physical zero mode in the sector of the additional variables. Because of
this reason, it is not obvious to us how to construct the
corresponding Lagrangian formulation, when generalizing the above
construction to superstrings. \footnote {For issues on the covariant
quantization of the Green-Schwarz strings see e.g. [16,17].}

\vspace{0.5cm}

\section*{Appendix}

Our convention for antisymmetrization of indices are as follows
\begin{eqnarray*}
&& A_{[ab]}=\displaystyle \frac 12 (A_aB_b-A_bB_a),\\
&& A_{[abc]}=\displaystyle\frac 13 (A_{a[bc]}+A_{b[ca]}+A_{c[ab]}),\\
&& A_{[abcd]}=\displaystyle\frac 14 (A_{a[bcd]}-A_{b[cda]}+
A_{c[dab]}-A_{d[abc]}),\\
&& A_{[abcde]}=\displaystyle\frac 15 (A_{a[bcde]}+A_{b[cdea]}+
A_{c[deab]}+A_{d[eabc]}+A_{e[abcd]}),\\
&& A_{[abcdef]}=\displaystyle\frac 16 (A_{a[bcdef]}-A_{b[cdefa]}+
A_{c[defab]}-A_{d[efabc]}+A_{e[fabcd]}-\\
&&\qquad\qquad -A_{f[abcde]}),\end{eqnarray*}\\
and so on.

In particular, given two totally antisymmetric tensors $J_{ab}$ and
$J_{abcd}$ one has
\begin{eqnarray*}
&& J_{[ab}J_{cd]}= \displaystyle\frac
13 (J_{ab}J_{cd}+ J_{ac}J_{db}+J_{ad}J_{bc});\\
&& J_{[ab}J_{cdem]}=
\displaystyle\frac 1{15} (J_{ab}J_{cdem}+
J_{ac}J_{demb}+J_{ad}J_{embc}+J_{ae}J_{mbcd}+J_{am}J_{bcde}-\\
&&\qquad -J_{de}J_{mabc}-J_{dm}J_{abce}-J_{bc}J_{dema}-J_{bd}J_{emac}-
J_{be}J_{macd}-J_{bm}J_{acde}+\\
&&\qquad +J_{em}J_{abcd}+J_{cd}J_{emab}+J_{ce}J_{mabd} +J_{cm}J_{abde});\\
&& J_{[abc\hat m}{J^m}_{d]}=\displaystyle\frac 14 (J_{abcm}{J^m}_d-
J_{dabm}{J^m}_c+J_{cdam}{J^m}_b-J_{bcdm}{J^m}_a).
\end{eqnarray*}

\end{document}